\documentclass[prl,aps,twocolumn,showpacs]{revtex4}
\usepackage{amsmath}
\usepackage{amsfonts}
\usepackage{amssymb}
\usepackage{graphicx}
\usepackage{float}
\usepackage{color}

\newcommand{\s}{\sigma}
\newcommand{\ltop}{h^\text{reg}}
\newcommand{\lbot}{h^\text{act}}
\newcommand{\lin}{m}
\newcommand{\lout}{\ell}
\newcommand{\ud}{\,\mathrm{d}}
\newcommand{\beq}{\begin{equation}}
\newcommand{\eeq}{\end{equation}}

\begin{document}

\title{Evolution of sparsity and modularity in a model of protein allostery}

\author{Mathieu Hemery} 
\affiliation{CNRS, LIPhy, F-38000 Grenoble, France}
\affiliation{Univ.~Grenoble Alpes, LIPhy, F-38000 Grenoble, France}

\author{Olivier Rivoire} 
\affiliation{CNRS, LIPhy, F-38000 Grenoble, France}
\affiliation{Univ.~Grenoble Alpes, LIPhy, F-38000 Grenoble, France}

\begin{abstract}
The sequence of a protein is not only constrained by its physical and biochemical properties under current selection, but also by features of its past evolutionary history. Understanding the extent and the form that these evolutionary constraints may take is important to interpret the information in protein sequences. To study this problem, we introduce a simple but physical model of protein evolution where selection targets allostery, the functional coupling of distal sites on protein surfaces. This model shows how the geometrical organization of couplings between amino acids within a protein structure can depend crucially on its evolutionary history. In particular, two scenarios are found to generate a spatial concentration of functional constraints: high mutation rates and fluctuating selective pressures. This second scenario offers a plausible explanation for the high tolerance of natural proteins to mutations and for the spatial organization of their least tolerant amino acids, as revealed by sequence analyses and mutagenesis experiments. It also implies a faculty to adapt to new selective pressures that is consistent with observations. Besides, the model illustrates how several independent functional modules may emerge within a same protein structure, depending on the nature of past environmental fluctuations. Our model thus relates the evolutionary history and evolutionary potential of proteins to the geometry of their functional constraints, with implications for decoding and engineering protein sequences.
\end{abstract}

\maketitle

Proteins are well known to be highly tolerant to mutations~\cite{Bowie:1990wk}. Statistical analyses of protein sequences~\cite{Lockless:1999uf} and saturated mutagenesis experiments~\cite{McLaughlinJr:2012hw} are now revealing the spatial architecture of this robustness: in several proteins, the amino acids most essential to the function are organized in small, structurally connected clusters of interacting and coevolving residues, called protein sectors~\cite{Halabi:2009jca}. For instance, in PDZ domains, a family of small interaction domains, a sector connects the ligand binding pocket to an opposite surface site~\cite{Lockless:1999uf,McLaughlinJr:2012hw}, which regulates allosterically the active site in at least one member of the family~\cite{Peterson:2004tl}. Similar sectors have been found and experimentally investigated in other protein families, which also consist of small structurally connected subsets of residues and, in many cases, mediate allostery~\cite{Suel:2003ca,Smock:2010bra,Reynolds:2011gs}. Furthermore, several quasi-independent sectors have been found to co-exist within a same protein domain~\cite{Halabi:2009jca}.

This spatial concentration of functional constraints within a protein structure is presently unexplained. It may be inherent to the physical properties of proteins, including the functional properties for which they were selected. For instance, when the function involves binding to a ligand, the residues structurally closer to the ligand may be expected to be functionally more important. We shall show, however, that such structural heterogeneities are not needed to explain a spatial concentration of functional constraints within a protein structure. To this end, we introduce below a simple mathematical model in which all "residues" are a priori equivalent, but where a sparse sector can nevertheless arise as a consequence of fluctuations during the evolutionary process.

The role of evolutionary history in shaping biological organizations has been discussed previously in relation to modularity, the generic decomposition of biological networks into subnetworks~\cite{Hartwell:1999wt}, of which the presence of several independent sectors within a same protein structure is an illustration~\cite{Halabi:2009jca}. Explanations for the origin of modularity broadly fall in two classes~\cite{Wagner:2007ti}: first, those based on the combinatorial properties of the process generating new variations, e.g., gene duplications and recombinations~\cite{Sole:2003tp}, and, second, those invoking the history of selective pressures, notably the particular structure of environmental fluctuations~\cite{Kashtan:2005wv}. In proteins, combinatorial reassortment may thus explain multi-domain architectures, and selection the decomposition of a domain into sectors that are intermingled along the sequence~\cite{Halabi:2009jca}. In general, however, the variational and selective factors are non-exclusive, and may contribute jointly to the emergence of modules~\cite{Sun:2007un}. Besides the question of their origin, the implications of modular architectures for future evolution have also been extensively studied in terms of resilience to mutations, or "robustness", and in terms of faculty to adapt, or "evolvability"~\cite{Wagner2005}, two properties found to have a complex relationship~\cite{Wagner:1996ut,Ancel:2000uz,Draghi:2010bj}.

The presence of a single sector in the network of interacting residues forming the structure of a protein corresponds to a degenerate form of modularity, better referred to as "sparsity"~\cite{Leclerc:2008ce}. Here, we demonstrate in the context of a physical model of protein evolution how sparsity generically emerges in the form of a spatial concentration of functional constraints from fluctuations during the evolutionary process. These fluctuations may involve variational or selective factors, and may promote robustness and/or evolvability to varying degrees. The phenomenon that we describe is more elementary than the evolution of modularity, which arises when the fluctuations have some additional structure, in relation to the structure of the function itself.

\section*{A model for the evolution of allostery}

To illustrate the role of evolutionary history in a context where structural heterogeneities are minimized, we introduce a model defined on a regular structure, and consider an allosteric property, which may in principle involve the entire structure. In the spirit of previous theoretical studies of protein evolution~\cite{Chan:2002vp}, we present this model in the generic framework of spin glasses~\cite{Bryngelson:1987uu}, but the Gaussian spin glass~\cite{Berlin:1952uo} that we analyze more specifically is also closely linked to models of elastic networks~\cite{Bahar:1997uy}. 

We may derive our model starting from a general expression for the energy of a protein,
\beq
\begin{split}
E=&-\sum_iK_0(a_i,\s_i,\varepsilon(r_i))-\sum_iK_1(a_i,a_{i+1},\s_i,\s_{i+1})\\
&-\sum_{i,j}K_2(a_i,a_j,r_i,r_j,\s_i,\s_j),
\end{split}
\eeq
where $a_i$ indicates the amino acid at position $i$ along the chain, $r_i$ its mean position, say of its alpha carbon, and $\s_i$ its physical state, say its orientation and fluctuations around the mean position. The term $K_0(a_i,\s_i,\varepsilon(r_i))$ represents an interaction energy between the amino acid $a_i$ and its local environment $\varepsilon(r_i)$, which may include the solvent and/or the amino acids of another protein, $K_1(a_i,a_{i+1},\s_i,\s_{i+1})$ represents the bonding energy between successive amino acids along the chain, and $K_2(a_i,a_j,r_i,r_j,\s_i,\s_j)$ the interactions of residues far apart along the chain but brought together upon folding.

While others have studied the incidence of evolutionary parameters on protein structure and stability~\cite{Taverna:2002jh,Tiana:2004kn}, we focus here on the evolution of amino-acid specific variables in the context of a fixed structure, to analyze the structural organization of functional constraints in members of a protein family sharing a common fold. We thus fix the $r_i$ at the nodes of a lattice, where only nearest neighbors have non-zero interactions. For simplicity, and to minimize structural heterogeneities, we also ignore the distinction between bond and non-bond energies, so that
\beq
E=-\sum_{\langle i,j\rangle}K(a_i,a_j,\s_i,\s_j)-\sum_iK_0(a_i,\s_i,\varepsilon(r_i)),
\eeq
where $\langle i,j\rangle$ indicates neighboring sites on the lattice. 

We further assume that the $\s_i$ take real values and that $K$ has the form $K(a_i,a_j,\s_i,\s_j)=J(a_i,a_j)\s_i\s_j$. Similarly, we assume that the environment around $i$ is represented by a real number $h_i$ with $K_0$ of the form $K_0(a_i,\s_i,\varepsilon(r_i))=h_i\s_i$ (more generally, $h_i$ could depend on $a_i$). We thus arrive at an energy of the form
\beq\label{eq:hamiltonian}
E\left(\s|a,h\right)=-\sum_{\langle i,j\rangle}J(a_i,a_j)\s_i\s_j-\sum_ih_i\s_i,
\eeq
which is formally the energy of a spin glass~\cite{SpinGlasses}, where spins $\s_i$ interact in the context of given (quenched) couplings $J(a_i,a_j)$ and fields $h_i$.

These simplifications are drastic but retain the essential relationships between the variables of the problems: the amino acid $a_i$, which may be subject to evolution, the environmental variables $h_i$, which may vary with time, and the physical variables $\s_i$, which are subject to short-range interactions constrained by the overall structure and are dependent on the identity of the amino acids and on the environment.

In this framework, we can apply the standard approach of statistical mechanics and sum over the internal degrees freedom $\s_i$ to compute from $E\left(\s|a,h\right)$ a free-energy $F\left(a,h\right)$. We can thus define a free energy of binding with an external ligand: the presence of a ligand corresponds indeed to a field $h'$ differing from the field $h$ in its absence, so a binding free energy is obtained as the difference $F\left(a,h'\right)-F\left(a,h\right)$. On the other hand, mutations induce a difference of free energy of the form $F\left(a',h\right)-F\left(a,h\right)$.

\begin{figure}
\begin{center}
\includegraphics[width=\linewidth]{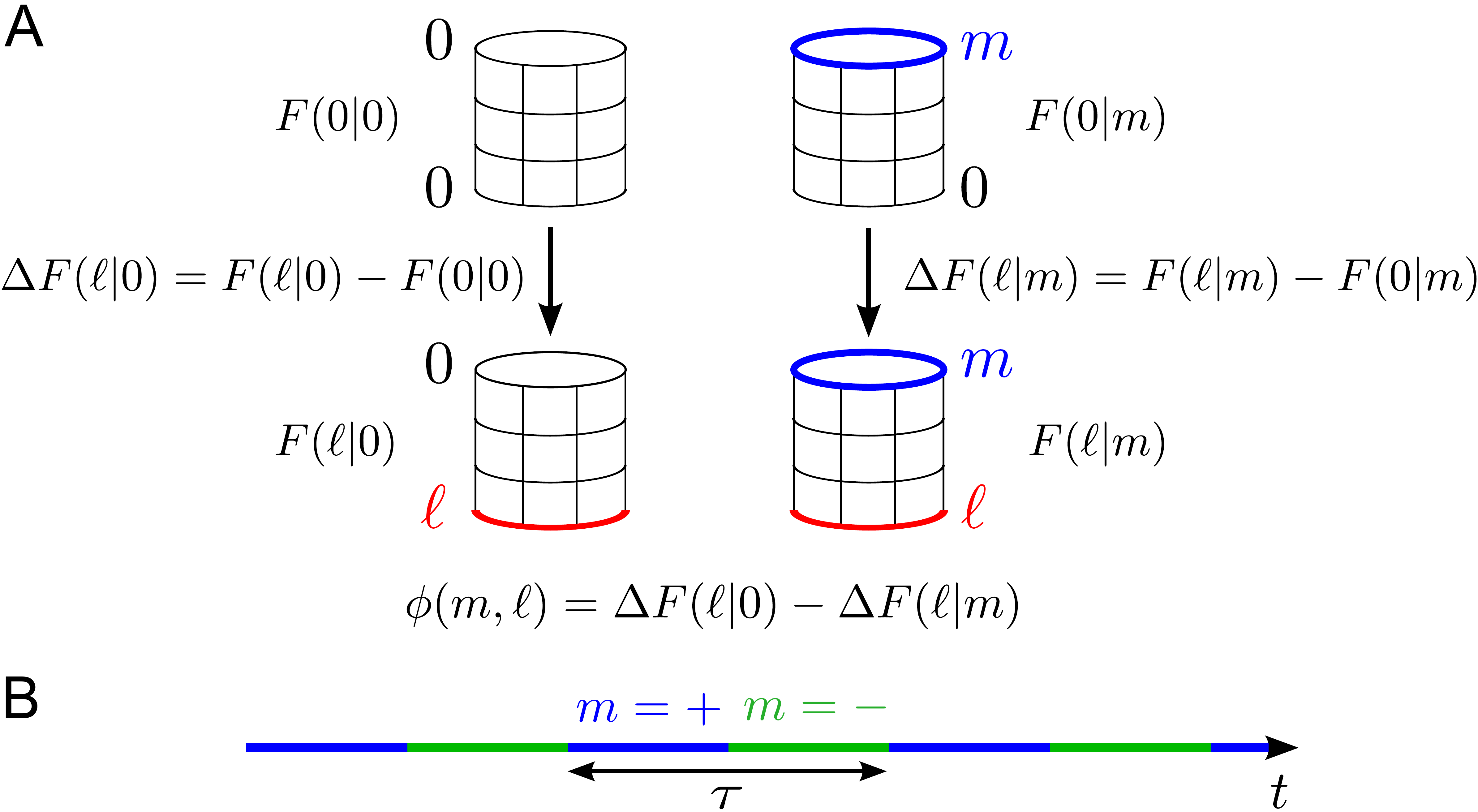}
\caption{{\bf A.\ }A model of protein allostery is defined as a spin glass on a cylindrical lattice. In this model, spins $\s_i$ are on the nodes to represent the physical states of residues, and couplings $J_{ij}$ are on the edges to represent interactions between residues, subject to evolution. Interactions with a modulator $m$ and/or a ligand $\ell$ are modeled by fields $h_i$ applied to the sites $i$ at the open ends of the cylinder. Allostery is quantified by $\phi(m,\ell)$, the difference between the binding free energy of $\ell$ in the presence of $m$, $\Delta F(\ell|m)$, and in its absence, $\Delta F(\ell|0)$.  {\bf B.\ }The sequences of the modulator and ligand are defined by the values and signs of the fields $h_i$, with $h_i=0$ representing an interaction with the solvent. Evolution is performed over a population of systems with selection for allosteric efficiency and with mutations affecting the couplings $J_{ij}$ at a rate $\mu$. A given generation is selected for allostery with given $\ell,m$ but when the environment fluctuates, different generations may experience different $\ell,m$. By symmetry, varying $\ell$ or $m$ is equivalent and we fix the sequence of the ligand to $h_i=+1$ for all $i$ at the bottom of the cylinder when $\ell$ is present, and vary only the sequence of the modulator every $\tau/2$ generations, between $h_i=+1$ for $i$ at the top ($m=+$) and $h_i=-1$ ($m=-$). Modulators or ligands with non-uniform sequences can be also considered as in Fig.\ S\ref{fig:ligands}. \label{fig:scheme}}
\end{center}
\end{figure}

Here, we consider a cylindrical lattice where two different ligands can bind at the two opposite open ends (Fig.\ \ref{fig:scheme}A). This allows us to quantify the preferential binding of one of the ligands in presence of the other, which corresponds to an allosteric regulation~\cite{Monod:1963jy}. Following the terminology used for allosteric proteins, we call "regulatory site" (abbreviated in 'reg') the upper end of the cylinder, "modulator" the ligand binding to it, "active site" ('act') its lower end, and "endogenous ligand" the ligand binding to it. Taking the interaction of site $i$ with the solvent to correspond to $h_i=0$, we thus have 
\beq\label{eq:hamiltonian}
E\left(\s|a,h\right)=-\sum_{\langle i,j\rangle}J(a_i,a_j)\s_i\s_j-\sum_{i\in\text{reg}} \ltop_i\s_i-\sum_{i\in\text{act}} \lbot_i\s_i,
\eeq
where $\ltop_i=m_i$ in the presence of a modulator characterized by the vector $m_i$ ($i\in\text{reg}$), $\ltop_i=0$ in its absence, and $\lbot_i=\ell_i$ in the presence of an endogenous ligand characterized by the vector $\ell_i$ ($i\in\text{act}$), $\lbot_i=0$ in its absence.

We define allostery as a more favorable interaction with a ligand $\lout$ in the presence of a modulator $\lin$. It is quantified thermodynamically in terms of free energy differences~\cite{Leff:1995wd}, as
\beq\label{eq:phi}
\phi(J|\lin,\lout)=\Delta F(\ell|0)-\Delta F(\ell|m)
\eeq
where $\Delta F(\ell|0)\equiv F(\ell|0)-F(0|0)$ represents the binding free energy of $\ell$ in absence of $m$, and $\Delta F(\ell|m)\equiv F(\ell|m)-F(0|m)$ in its presence, as illustrated in Fig.\ \ref{fig:scheme}A (the amino acids are fixed in these expressions).

We simulate an evolutionary dynamics by a standard genetic algorithm~\cite{Mitchell}, whereby a population of $P$ systems undergoes repeated cycles of selection, reproduction and mutation. Selection and reproduction are based on allosteric efficiency, as defined by Eq.~\eqref{eq:phi}, with systems with larger $\phi(J)$ generating more offsprings. Mutations of the amino acids correspond to changes of the couplings; for simplicity, instead of introducing an arbitrary matrix $J(a,b)$, we assume that a mutation randomly change the value of a single coupling $J_{ij}=J(a_i,a_j)$ at a rate $\mu$ per generation, independently of the other couplings (Materials and methods); we verified, however, that explicitly mutating amino acids at the level of sites, which affect simultaneously several couplings, lead to similar results.

Numerical simulations of evolutionary dynamics are generally limited by the computational cost of estimating the fitness of each individual. Here, four free energies of spin glass models on finite-dimensional lattices are involved; their computations would be very demanding if considering Ising spins $\s_i=\pm 1$~\cite{Barahona:1982ui}, but by considering a Gaussian model~\cite{Berlin:1952uo}, for which $\phi(J)$ can be expressed analytically for any couplings $J_{ij}$ and any geometry of the lattice, the computations are reduced to the inversion of a matrix (Materials and methods). In this Gaussian model, the spins $\s_i$ take arbitrary real values, but the couplings $J_{ij}$ need to be bounded: we thus mutate the couplings by drawing them uniformly in $[-1,+1]$. This Gaussian model may also be viewed as an elastic network model~\cite{Bahar:1997uy} with a single degree of freedom per site and non-uniform "spring constants".

\section*{Evolutionary concentration of functional constraints}

\begin{figure}
\begin{center}
\includegraphics[width=\linewidth]{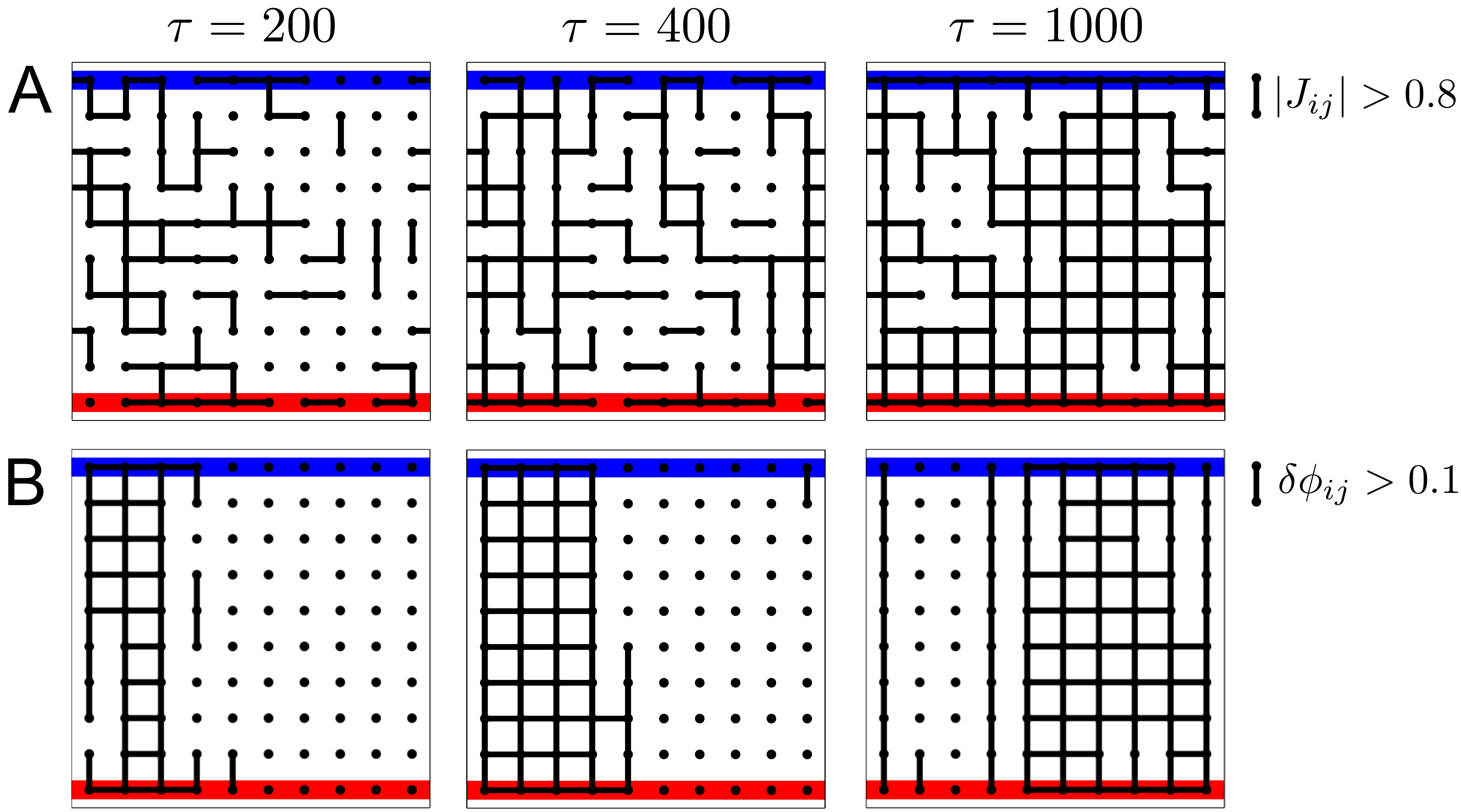}
\caption{Examples of systems obtained from an evolutionary dynamics with mutation rate $\mu=5.10^{-5}$ and different periods $\tau$ of fluctuations of selective pressure ($\tau=200, 400, 1000$). {\bf A.\ } Couplings $J_{ij}$ with large absolute values, $|J_{ij}|>0.8$. {\bf B.\ } Couplings $J_{ij}$ inducing a large loss in allosteric efficiency when mutated, $\delta\phi_{ij}>0.1$ (see Fig.\ S\ref{fig:cutoffs} for other values of the cut-offs). A third approach, based on coevolution, can also reveal the same concentration of functional constraints (Fig.\ S\ref{fig:coevo}). The figures display the fittest individual in a population of $P=500$ individuals prior to a change of environment.
\label{fig:grids}}
\end{center}
\end{figure}

The outcome of the evolutionary dynamics is contingent on the series of modulator and ligand sequences that the successive generations encounter (Fig.\ \ref{fig:scheme}B). When these sequences are constant over time, say $\lin=(+1,\dots,+1)$ and $\lout=(+1,\dots,+1)$ at all time, systems evolve maximal couplings $|J_{ij}|\simeq 1$ at all sites. This implementation of the couplings optimizes the allosteric efficiency $\phi$ and epitomizes an absence of sparsity. Repeating the same simulations with a modulator that alternates with period $\tau$ between two sequences, $m=(+1,\dots,+1)$ and $m=(-1,\dots,-1)$, yields a qualitatively different outcome: the smaller $\tau$ is, the fewer are the large couplings, as illustrated in Fig.\ \ref{fig:grids}A.

Allostery requires strong couplings, but not all strong couplings need to be functionally significant: if a strong coupling is defined by $|J_{ij}|>0.8$ as in Fig.\ \ref{fig:grids}A, we may indeed expect $\sim 20\%$ of strong couplings even in absence of any selection, only because the $J_{ij}$ are mutated to random values in $[-1,1]$ ($0.8$ is an arbitrary cut-off but other values lead to a similar conclusion, see Fig.\ S\ref{fig:cutoffs}). As a more relevant measure of "functional significance", we may consider instead the "fitness value" $\delta\phi_{ij}$ of $J_{ij}$, defined as the maximal cost that a mutation of $J_{ij}$ can cause to the allosteric efficiency $\phi$ (Material and methods). We thus identify functionally significant couplings $J_{ij}$ by $\delta\phi_{ij}>\epsilon$, with for instance $\epsilon=0.1$ in Fig.\ \ref{fig:grids}B (see Fig.\ S\ref{fig:cutoffs} for other values of $\epsilon$). This criterion, closer to what has been experimentally measured~\cite{McLaughlinJr:2012hw}, reveals distinctly the presence of a connected subset of functional couplings joining the regulatory and active sites (Fig.\ \ref{fig:grids}B).

This subset of functionally significant couplings, which breaks the rotational invariance of the cylinder and whose location varies from simulation to simulation, displays several features reminiscent of protein sectors observed in natural proteins~\cite{Lockless:1999uf,Halabi:2009jca}: (i) it is overall structurally connected (Fig.\ \ref{fig:grids}B); (ii) it has a hierarchical organization: less significant couplings are peripheral to more significant ones, as shown by varying the value of the cut-off $\epsilon$ defining functional significance (Fig.\ S\ref{fig:cutoffs}); (iii) it is evolutionarily conserved: its location is stable over multiple periods along a given evolutionary trajectory (Fig.\ S\ref{fig:stability}); (iv) its couplings are coevolving, as shown by a statistical analysis of a 'multiple sequence alignment' obtained from independent evolutionary trajectories with a common origin (Fig.\ S\ref{fig:coevo}).

\begin{figure}
\begin{center}
\includegraphics[width=.9\linewidth]{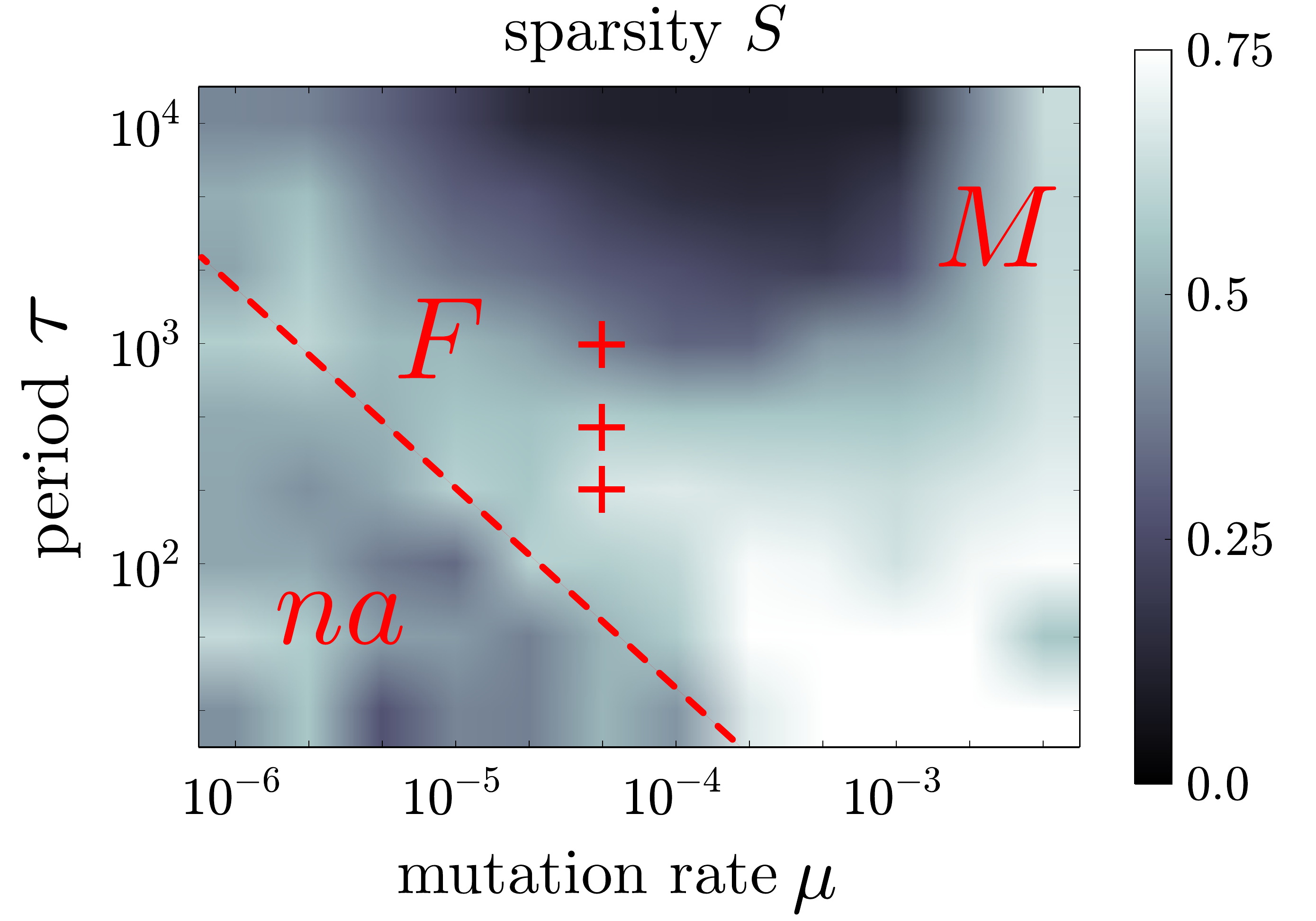}
\caption{Sparsity of evolved proteins as a function of the mutation rate $\mu$ and the period $\tau$ of fluctuating selective pressures. Sparsity is defined as the fraction of couplings with $\delta\phi_{ij}<0.1$ (non-represented couplings in Fig.\ \ref{fig:grids}B). Below the dotted line, the environmental fluctuations are too fast for the population to follow them, and the systems are non-adapted (region $na$, see also Fig.\ S\ref{fig:unfit}). Sparse systems are found in two ranges of parameters: for intermediate values of $\mu\tau$, where they are driven by fluctuating selective pressures (region $F$, including the three systems of Fig.\ \ref{fig:grids} indicated by crosses), and for high values of $\mu$, where they are driven by a large mutational load (region $M$).\label{fig:parameters}}
\end{center}
\end{figure}

As indicated by Fig.\ \ref{fig:grids}B, the smaller the period $\tau$ of the fluctuations in selective pressure, the smaller the sector. The temporal structure of past selective objectives is thus encoded geometrically in the couplings. More precisely, we may define the sparsity of a system as the fraction $S$ of its couplings $J_{ij}$ with $\delta\phi_{ij}<0.1$ (the fraction of non-represented couplings in Fig.\ \ref{fig:grids}B). This measure of sparsity, is represented in Fig.\ \ref{fig:parameters} as a function of the mutation rate $\mu$ and of the period $\tau$. At not too high mutation rates (see below), it scales with $\mu\tau$, the number of mutations per period; more precisely, it scales with $\mu\tau P$, the total number of mutations in a population of size $P$ (Fig.\ S\ref{fig:scaling}). 

While sparsity arises at the expense of instantaneous fitness, here defined by the allosteric efficiency $\phi$ (Fig.\ S\ref{fig:unfit}), it favors the "evolvability"~\cite{Wagner:1996ut} of the population, which can be quantified as the fraction of random mutations conferring a noticeable fitness advantage following a change of selective objective (Suppl. Appendix, Figs.\ S\ref{fig:slices_tau}C-S\ref{fig:slices_mu}C). The evolution of sparsity also implies an increased mutational "robustness"~\cite{vanNimwegen:1999ut}, defined as the fraction of mutations that do not affect noticeably the fitness (Suppl. Appendix, Figs.\ S\ref{fig:slices_tau}B-S\ref{fig:slices_mu}B). 

The period $\tau$ is not the only feature of the environmental fluctuations that affects the size of a sector: so does the diversity of these fluctuations. For a given $\tau$, the sparsity thus decreases linearly with the sequence similarity between the two alternating modulators (Fig.\ S\ref{fig:ligands}A). But while the similarity between successive modulators is determining, their exact sequence is not: replacing the sequences $m=(+1,\dots,+1)$ and $m=(-1,\dots,-1)$ by arbitrary sequences of $\pm 1$, or even imposing new randomly chosen modulators at each period, does not affect significantly the outcome (Fig. S\ref{fig:ligands}B). This observation illustrates a capacity of "generalization"~\cite{Parter:2008bb}: the sparse systems, which are more prompt to re-adapt to a modulator previously encountered in their history, are as prompt to adapt to a modulator never encountered. 

Another factor besides the fluctuations of selective pressures can induce a sector: a large mutational load. While for small mutations rates $\mu$ the sparsity is controlled by the dimension-less parameters $\mu\tau$, for large mutation rates it is controlled by $\mu$ nearly independently of $\tau$ (Figs.\ \ref{fig:parameters} and S\ref{fig:slices_mu}A). The critical value of the mutation rate, $\mu_c\sim N^{-1}$, corresponds to the "error-threshold" for a system of size $N$ (here the number of links $ij$), i.e., to the maximal mutation rate at which a system of this size can faithfully replicate~\cite{Hypercycle}. For $\mu>\mu_c$,  the systems thus evolve a sector of size $\sim (\mu N)^{-1}$, which is the largest size allowed by the mutational load. The sparse systems obtained at high $\mu$ are more robust and less evolvable than the sparse systems obtained in fluctuating environments at low $\mu$, but nevertheless more evolvable than the non-sparse systems obtained at lower values of $\mu$ (Figs.\ S\ref{fig:slices_tau}-S\ref{fig:slices_mu}). A difference is also apparent at the population level: the variance in the population of the couplings outside the sectors is low at low $\mu$ and large at large $\mu$.

\section*{Localization in sequence space}

\begin{figure}
\begin{center}
\includegraphics[width=.9\linewidth]{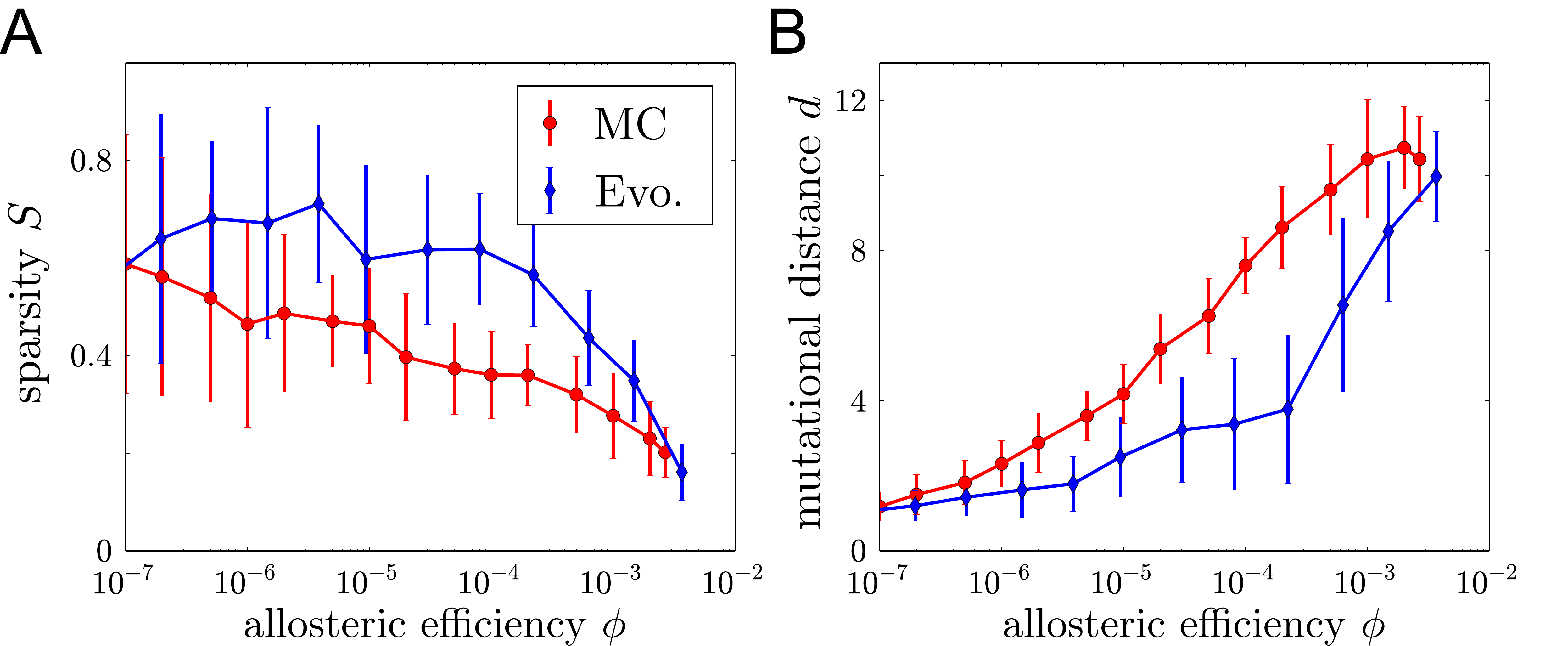}
\caption{{\bf A.\ }Sparsity as a function of allosteric efficiency (fitness) for evolved systems (in blue, various $\mu<10^{-2}$ and $\tau>10$) and for typical systems with same fitness (in red), obtained by Monte Carlo sampling (Materials and methods). {\bf B.\ }Distance to a system with different function (allostery induced by a modulator $m'$ different from the one $m$ for which the system was last selected), measured by the minimal number of beneficial mutations needed to reach an equivalent allosteric efficiency after the change $m\to m'$, for evolved systems (blue) and typical systems (red). Systems evolved in a fluctuating environment are thus atypical amongst systems with equivalent fitness value for being sparser and closer to solutions to new selective challenges.\label{fig:localization}}
\end{center}
\end{figure}

Functional proteins represent only a tenuous subset of all potential proteins~\cite{Keefe:2001fa}. In our model, we find that within this subset, proteins with a sparse sector are themselves rare: typical systems with a given fitness $\phi$ are significantly less sparse than systems with the same fitness but resulting from an evolution in fluctuating environments (Fig.\ \ref{fig:localization}A). This observation implies that sparsity in the evolved systems is not just a consequence of the fitness being curbed by the environmental fluctuations. The sparse systems are, besides, not distributed randomly in sequence space, but localized in evolvable regions of this space: they are at shorter mutational distance to solutions to alternative selective pressures (Fig.\ \ref{fig:localization}B). This phenomenon of localization is generic and has been  illustrated previously in the context of fitness landscapes defined on small, schematic sequence spaces~\cite{Meyers:2005ep,Kussell:2006vp}.

Our model, however, displays two features absent from simpler models. First, it relates the topology of the fitness landscape, defined in sequence space, to the geometry of the functional constraints, defined in real space: gradients in fitness thus map to sector positions, where adaptive mutations occur, while plateaux in fitness map to non-sector positions, where neutral mutations occur. Second, as typical to high-dimensional spaces, the results are partly non-intuitive: a system localized between two alternating fitness peaks is thus ipso facto localized near a large family of related fitness peaks (Fig.\ \ref{fig:localization}B), a feature that underlies the faculty of generalization~\cite{Parter:2008bb}, or "promiscuity"~\cite{Tawfik:2010jp}, previously noted.

\section*{Modularity}

The concentration of functional constraints may take different geometrical forms depending on the structure of the evolutionary fluctuations. In particular, distinct quasi-independent sectors may evolve instead of a single connected sector. A combinatorial process for generating new variations, involving for instance gene duplications, recombination events and/or horizontal transfers, has for instance been shown to produce modular organizations~\cite{Sole:2003tp}. Such combinatorial variations may explain the modular organization of proteins into domains, which are subsequences of consecutive amino acids, but cannot easily account for the presence of multiple quasi-independent sectors distributed along the sequence of a single domain~\cite{Halabi:2009jca}. A scenario implicating modularly varying selective pressures provides an alternative explanation, as previously illustrated in a range of different models~\cite{Kashtan:2005wv,Parter:2008bb,Kashtan:2009hu}. 

\begin{figure}
\begin{center}
\includegraphics[width=\linewidth]{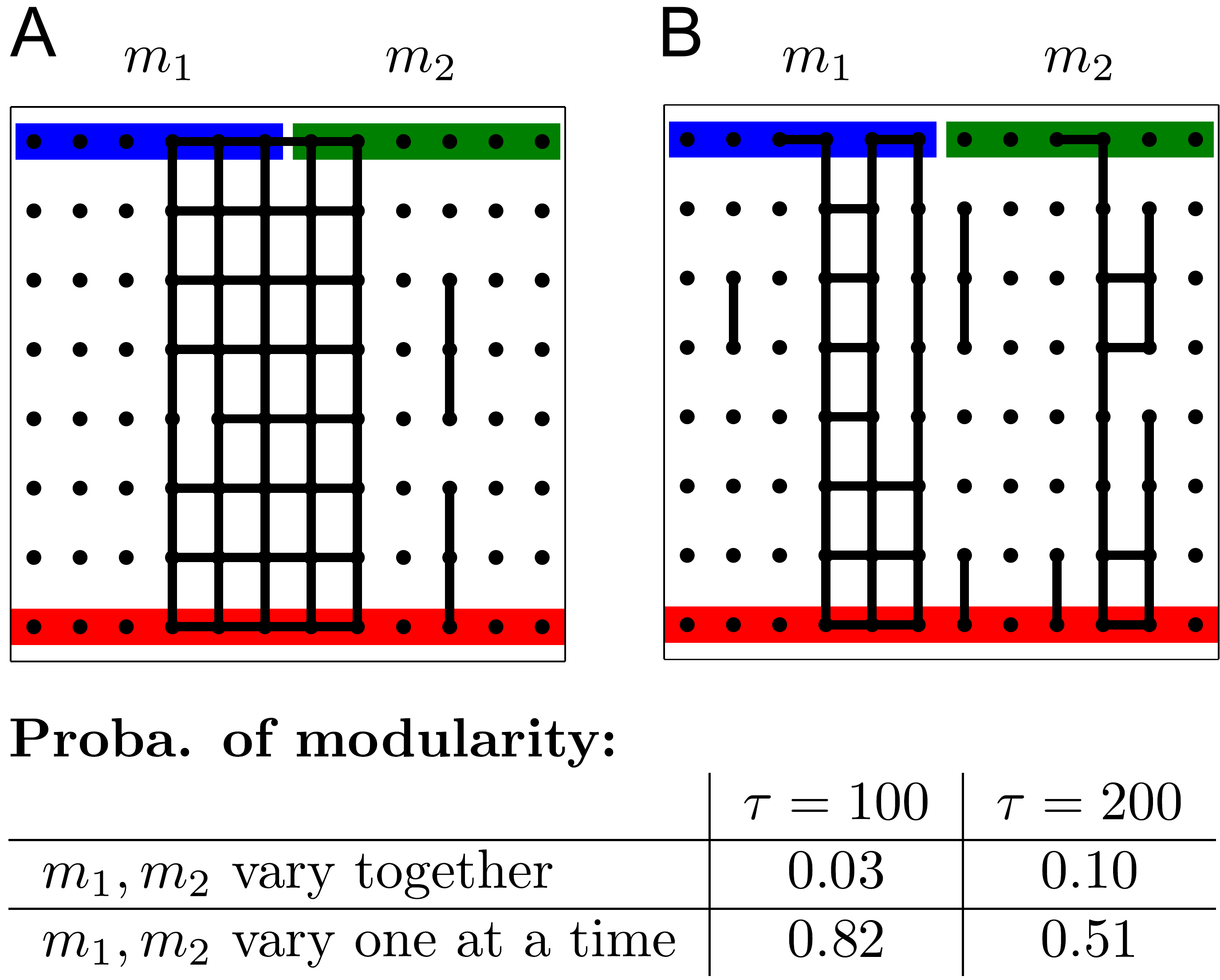}
\caption{Typical outcomes for a variant of the model where the regulatory site is partitioned into two sub-sites, each associated with an independently varying modulator, and where selection is on allostery in the presence of at least one of the two modulators $m_1$ and/or $m_2$. {\bf A.\ } A non-modular system, with a single sector localized at the interface between the two regulatory sites. {\bf B.\ } A modular system, with two distinct sectors. These two systems are the (stable) outcomes of two distinct evolutionary trajectories with same evolutionary parameters ($\tau=100$, $\mu=5.10^{-4}$). As for the location of the sector in Fig.\ \ref{fig:grids}, the difference stems from stochastic effects. The table indicates the probability to obtain a modular system for two values of the period $\tau$ and for $m_1,m_2$ varying either simultaneously or consecutively.\label{fig:modularity}}
\end{center}
\end{figure}

Consistently with these past works, we find that a modularly varying environment favors the emergence of two distinct sectors in an extension of our model where allostery involves two possible modulators. In this model, the two modulators $m_1$, $m_2$ can bind at two distinct regulatory sites (Fig.\ \ref{fig:modularity}), and selection is for preferential binding of the ligand $\ell$ in the presence of at least one of them (Material and methods). When both the sequences of $m_1$ and $m_2$ fluctuate in time, evolution stochastically generates one of two possible outcomes: systems with a single sector, as in Fig.\ \ref{fig:modularity}A, or with two separate sectors, as in Fig.\ \ref{fig:modularity}B. The probability to obtain two sectors depends on the structure of the fluctuations (besides the size of the structure): it is significantly larger when $m_1$ and $m_2$ change modularly, i.e., one at a time, compared to when they change non-modularly, i.e., simultaneously (Table in Fig.\ \ref{fig:modularity}).

We note that, in contrast with previous models reporting similar effects~\cite{Kashtan:2005wv,Parter:2008bb,Kashtan:2009hu}, sparsity is not enforced in the definition of our fitness, but obtained as a result of evolution. We also find that a rugged fitness landscape is not necessary for modularity to emerge spontaneously~\cite{Sun:2007un}: in our model, solutions are indeed always accessible by hill-climbing with one-step mutations (Fig.\ \ref{fig:localization}B). 

\section*{Discussion}

Interpreting the information contained in the sequence of a protein does only require referring to its biophysical properties, but also to its evolutionary history. Our simple model of protein evolution thus demonstrates how a basic feature of proteins, the spatial organization of their residues least tolerant to mutations, may be controlled by past fluctuations of selective pressure or high mutation rates. Our conclusions are based on comparing scenarios that differ only in two evolutionary parameters, the period $\tau$ of environmental fluctuations and the mutation rate $\mu$. Since a structural concentration of functional constraints arises only for some values of these parameters, it is clearly not a necessary consequence of the definition of our model. 

We expect that comparable results hold for other systems where internal variables varying on a short time scale are similarly subject to short range interactions controlled by evolutionary and environmental variables varying on longer time scales. In less idealized systems, including natural proteins, several additional factors may, however, contribute to a concentration of functional constraints. 

Irregular structures thus typically contain preferred allosteric paths that tend to reinforce the concentration of functional constraints: with no unique shortest path between its two interfaces, the cylindrical structure allowed us to illustrate the role of evolutionary factors with minimal contribution from structural heterogeneities. Our approach, however, extend to other geometries (Fig.\ S\ref{fig:geometry}).

Similarly, our results are robust to variations in the implementation of the evolutionary dynamics, but alternative choices may reduce or enhance sparsity; for instance, a multiplicative mutational process, which is biased towards vanishing couplings, generically favors sparsity over an additive process (Fig.\ S\ref{fig:mutations}) \cite{Friedlander:2013jr}. In our model, all coupling values are a priori equiprobable, showing not only that a mutational bias is not required, but also that sparsity of functional constraints, as reported by the fitness cost of mutations, does not equate sparsity of the underlying physical couplings (Fig.\ \ref{fig:grids}). 

Sparsity may also be favored by factors limiting the efficiency of selection. The typically non-linear relationship between the biophysical properties of a protein and the reproductive rate (fitness) of organisms may thus make the contribution of all couplings unnecessary. Finite population size effects, which our genetic algorithm minimizes, also generically exclude a complete "optimization" of the couplings.

Our model represents an ideal case where, under constant environment, all the couplings may be equivalently involved in the function (the only a priori difference being between vertical and horizontal couplings). In the generic case where a uniform distribution of the couplings is intrinsically non-optimal, evolutionary fluctuations may, nevertheless, control the degree of concentration of functional constraints if they are sufficiently important.

Many extensions of our model are conceivable. Negative selection against undesired modulators and ligands may for example allow us to account for the specificity of the interactions. The assumption of a fixed geometry of interactions may also be relaxed to permit a joint treatment of folding and functional constraints, in line with previous studies based on similar simplified protein models~\cite{Hirst:1999tc,Williams:2001ua,Bloom:2004fs}. Extending our model to account for structural changes and kinetic effects may thus contribute to rationalize the diversity of mechanisms that evolved to cause allostery~\cite{Motlagh:2015kc}.

Our model is not intended to account quantitatively for the features of natural proteins. Nevertheless, given the typical size $N\sim 10^2$ and mutations rates $\mu\sim 10^{-9}$/bp/generation of current non-viral proteins, we may exclude a scenario based on high mutation rates for explaining the high tolerance of proteins to mutations. On the other hand, estimates of $\mu P$ based on silent genomic variations within species give $\mu P\sim 10^{-1}-10^{-3}$ for a range of organisms~\cite{Lynch06}, where $P$ represents an effective population size. This indicates that relevant time scales for a scenario based on fluctuating selective pressures are of the order of $\tau\sim (\mu P)^{-1}\sim10-1000$ generations; these estimations are crude but lend weight to the plausibility of this scenario. Differences of variability in past selective pressures may thus cause different proteins to have fundamentally different architectures of functional constraints.  

While our limited knowledge of past evolutionary history prevents us from testing quantitatively these ideas with natural proteins, progress in the field of directed evolution~\cite{Romero:2009kz,Esvelt:2011cv,Andrew12} may soon offer us a platform to investigate them experimentally.

\section*{Materials and methods}

{\bf Allosteric efficiency --} A Gaussian spin-glass model is defined at inverse temperature $\beta$ by the partition function~\cite{Berlin:1952uo} 
 \beq\label{eq:Z}
Z\left(h|J\right)=\int \prod_{i}\frac{e^{-\s_{i}^{2}/2}}{\sqrt{2\pi}}\ud\s_i\ e^{-\beta H(\sigma|J,h)},
\eeq
where $H(\sigma|J,h)=-\frac{1}{2}\s^\top J\s-h^\top\s$ is the Hamiltonian of Eq.~\eqref{eq:hamiltonian}, with the geometry of the lattice defined by the non-zero elements of the matrix $J_{ij}$ (with $J_{ii}=0$ and $J_{ij}=J_{ji}$). The model is defined only at high temperature since the integral diverges for large $\beta$. If $c$ denotes the maximal connectivity of the lattice, it is sufficient to assume that $|\beta h_i|,|\beta J_{ij}|<1/c$, which, on a square lattice with $|h_i|$, $|J_{ij}|\leq 1$, we achieve by fixing $\beta=0.1$. Under this assumption, the partition function is obtained by performing the Gaussian integration:
\beq
Z\left(h|J\right) =(\text{det}\left(I-\beta J\right))^{-1/2}\exp\left(\frac{\beta^{2}}{2}h^\top\left(I-\beta J\right)^{-1}h\right),
\eeq
where $I$ represents the $M\times M$ identity matrix, $M$ being the number of nodes in the lattice. The free energy $F(h|J)=-\beta^{-1}\ln Z(h|J)$ has hence the form
\beq
F(h|J) =-\frac{1}{2}\beta\ h^\top\left(I-\beta J\right)^{-1}h + F(0|J),
\eeq
where $F(0|J)$ does not depend on $h$, leading to the following expression for $\phi(J|m,\ell)$, defined in Eq.~\eqref{eq:phi}:
\beq
\phi(J|m,\ell) =\beta\ m^\top[\left(I-\beta J\right)^{-1}]_{\text{reg},\text{act}}\ \ell
\eeq
where $[A]_{\text{reg},\text{act}}$ denotes a sub-matrix of $A_{ij}$ where $i$ is restricted to $i\in\text{reg}$ and $j$ to $j\in\text{act}$.\\

{\bf Evolutionary algorithm --} Simulations were performed over $5.10^4$ generations with populations of $P=500$ systems consisting of $10\times 10$ square lattices. At each generation, a system $k$ with couplings $J_{ij}^k$ is replicated $n_k$ times based on the value of $\phi_k=\phi(J^k|m,\ell)$, following the sigma-scaling rule~\cite{Mitchell}, $n_k=1+(\phi_k-\overline{\phi})/(2\s_\phi^2)$, where $\overline{\phi}$ and $\s_\phi^2$ are respectively the mean and variance of $\phi_k$ in the population. For each system, each coupling $J_{ij}$ has then a probability $\mu$ to be mutated to a random value in $[-1,1]$. See Suppl.\ Appendix for other replication and mutational rules.\\

{\bf Functionally significant couplings -- } To compare systems with different allosteric efficiencies, we define the relative fitness cost of a mutation $J_{ij}\to J_{ij}^*$ as $\delta\phi^*_{ij}(J)\equiv (\phi(J)-\phi(J^{(*)}))/\phi(J)$, where $J^{(*)}$ differs from $J$ by the value of $J_{ij}$. In Fig.\ \ref{fig:grids}B, we consider $\delta\phi_{ij}$, the highest fitness cost of a mutation at $ij$, estimated by comparing the fitness costs of $J^*_{ij}=0,\pm1$ and by retaining the highest one. A coupling is said to be functionally significant if $\delta\phi_{ij}>\epsilon$ with $\epsilon=0.1$ (other choices of this cut-off yield similar results, see Fig.\ S\ref{fig:cutoffs}).\\

{\bf Sparsity -- } The sparsity $S$ of a system is defined as the fraction of its couplings with $\delta\phi_{ij}<\epsilon$, using $\epsilon=0.1$ (other choices of this cut-off yield similar results).\\

{\bf Sampling of systems with given fitness -- } To sample typical systems with a given value of fitness $\phi^0$, we implemented a standard Monte Carlo sampling algorithm with $|\phi(J|h) - \phi^0|$ as energy function.\\

{\bf Distance to alternative solutions -- } In Fig.\ \ref{fig:localization}B, the distance to an alternative selective pressure $h'\neq h$ is estimated as the number of steps necessary for a hill-climbing algorithm, whereby a single coupling $J_{ij}$ can be changed at each step, to reach a fitness value $\phi(J'|h')$ at least equivalent to the initial value $\phi(J|h)$.\\

{\bf Evolution of modularity -- } In a variant of our model, the regulatory site is split into two consecutive segments where two modulators $m_1$ and $m_2$ can bind, corresponding formally to $m=(m_1,m_2)$. Requiring the binding of a ligand $\ell$ to be allosterically regulated by the presence of any of the two modulators (non-exclusive OR) corresponds to selecting with a fitness $\phi = \text{min}(\phi_1,\phi_2)$, where the allosteric efficiencies $\phi_1$ and $\phi_2$ are defined by Eq.\ \eqref{eq:phi} with, respectively, $m=(m_1,0)$ and $m=(0,m_2)$. For the statistics shown in the table of Fig.\ \ref{fig:modularity}, a system is considered as modular if removing the couplings below $m_1$ (by setting the couplings to $0$) leads to a $\phi_2$ within $80 \%$ of the original $\phi$ and removing those below $m_2$ to a $\phi_1$ within $80 \%$ of the original $\phi$ (this definition is consistent with a classification based on visual inspection of networks as shown in Fig.\ \ref{fig:modularity}; it is somewhat arbitrary but the trends shown in the table of Fig.\ \ref{fig:modularity} are not).\\

\begin{acknowledgments}
We thank A. Dawid, D. Hekstra, B. Houchmandzadeh, I. Junier,  S. Leibler, C. Nizak, K. Reynolds, A. Raman and R. Ranganathan for discussions and comments. This work was supported by ANR grant CoevolInterProt.
\end{acknowledgments}

\bibliographystyle{unsrt}

\appendix
\clearpage
\section*{SUPPLEMENTARY APPENDIX}
\setcounter{figure}{0}
\makeatletter
\makeatletter \renewcommand{\fnum@figure}
{\figurename~S\thefigure}
\makeatother

{\bf Genetic algorithm --} The results presented in the main text are obtained with the sigma-scaling procedure described in Materials and methods. This procedure ensures that the variance in the number of offsprings remains constant despite a decrease of variance in fitness. If taking the number of offsprings directly proportional to the fitness, the decrease over time of the fitness variance indeed renders selection ineffective (thus requiring longer simulations and/or larger populations). An alternative to sigma-scaling is an elite strategy whereby, at each generation, the top $x$ \% individuals with best fitness are duplicated while the bottom $x$ \% are eliminated. We verified that with $x=20$, this strategy gives results equivalent to the sigma-scaling procedure.\\

\begin{figure}[H]
\begin{center}
\includegraphics[width=\linewidth]{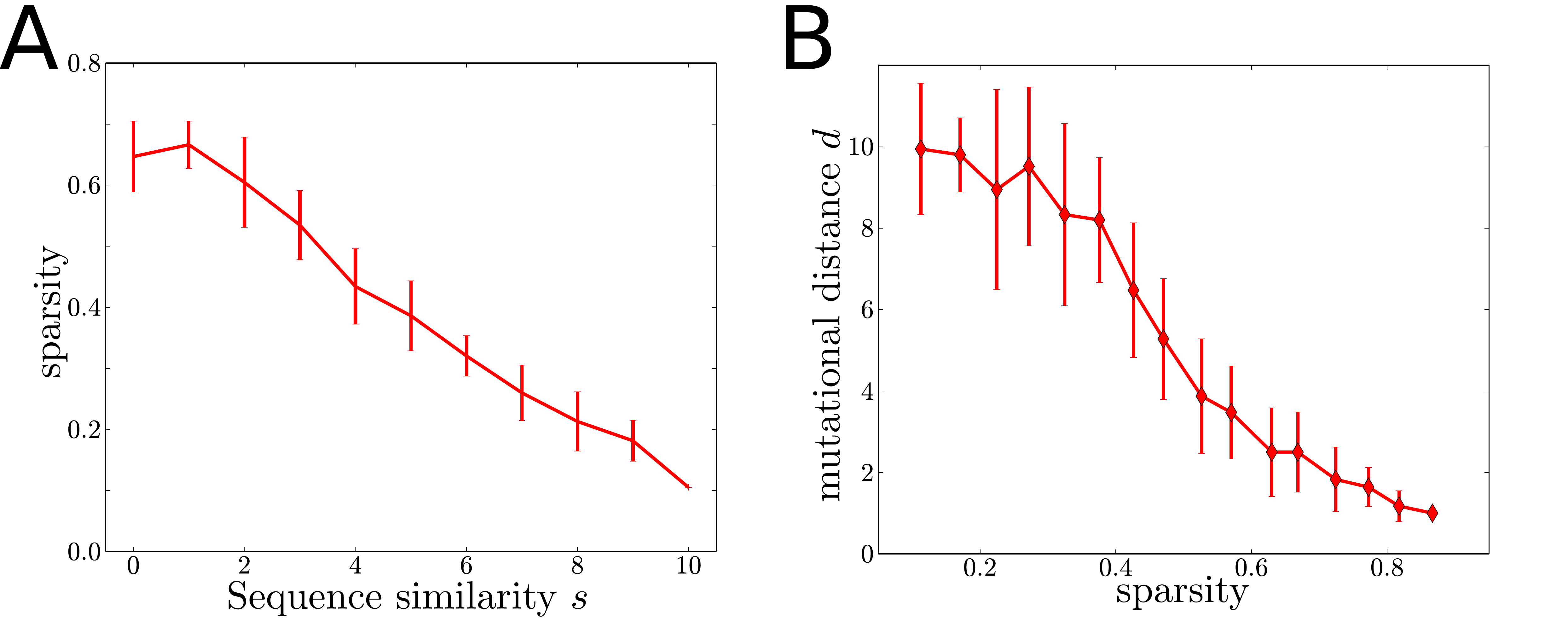}
\caption{
{\bf A.}~Sparsity as a function of the sequence similarity $s$ between the two alternative modulator sequences, for $\mu=10^{-5}$ and $\tau=100$ (mean and standard deviation over 10 simulations).
{\bf B.}~Minimal number of point mutations necessary to adapt to a new random modulator as a function of sparsity. The simulations are obtained with different values of $\tau \in [10,\ldots,5000]$ and $\mu \in [5.10^{-3},\ldots,2.10^{-6}]$, when they correspond to adapted populations ($\phi > 10^{-7}$); the results are averages over 5 random modulators.
\label{fig:ligands}}
\end{center}
\end{figure}

{\bf Different modulators --} In the main text, we present results when alternating between two opposite modulator sequences, $m^{(1)}=(+,\dots,+)$ and $m^{(2)}=(-,\dots,-)$. Any other choice of two opposite modulators with $m^{(1)}_i=-m^{(2)}_i$ for all $i\in {\rm act}$ gives identical results as a consequence of the "gauge invariance": $\s_i\mapsto-\s_i$ $\Leftrightarrow$ $J_{ij}\mapsto-J_{ij}$ $\forall j$. When alternating between two modulators $m^{(1)}$ and $m^{(2)}$ with $m_i^{(1)}=\pm 1$, $m_i^{(2)}=\pm 1$ and sequence similarity $s=\sum_{i}\delta(m^1_i,m^2_i)$, the sparsity is commensurate with this measure of similarity (Fig.\ S\ref{fig:ligands}A), thus interpolating between the case $s=10$, which is equivalent to a constant environment, and the case $s=0$, which corresponds to opposite modulators.

In Fig.\ \ref{fig:localization}B, we consider systems that evolved under an environment fluctuating between two opposite modulators and, for a system at the end of a period of constant environment, represent the minimal number of mutations necessary to achieve an equivalent allosteric efficiency in the presence of the other modulator. Fig.\ S\ref{fig:ligands}B shows that similar results are obtained when considering a random modulator not previously encountered.\\

\begin{figure}[H]
\begin{center}
\includegraphics[width=\linewidth]{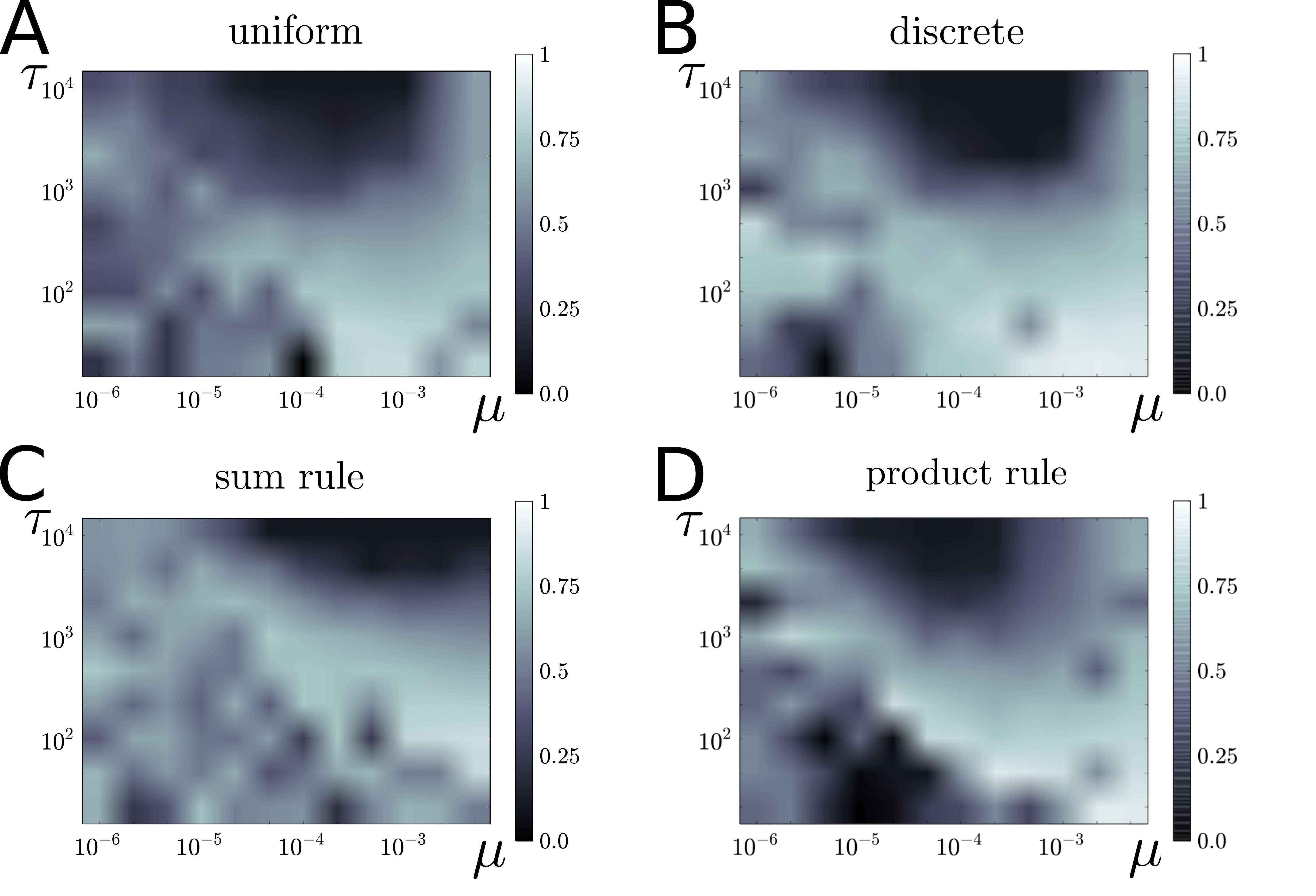}
\caption{Sparsity as a function of mutation rate $\mu$ and time scale $\tau$ of environmental changes for systems that evolved subject to different mutational processes: {\bf A.\ } As in Fig.\  \ref{fig:parameters}, the $J_{ij}$ are mutated to a random value uniformly distributed in $[-1,1]$, independently of their previous value.  {\bf B.\ }The $J_{ij}$ are drawn from a finite set of discrete values. {\bf C.\ }Sum-rule with variance $\sigma_s^2 = 0.2$. {\bf D.\ }Product-rule with variance $\sigma_p^2 = 1.6$. In each case and as in Fig.\ \ref{fig:parameters}, the left bottom corner corresponds to non-adapted populations.\label{fig:mutations}}
\end{center}
\end{figure}

{\bf Alternative mutational processes --}
The results presented in the main text are obtained with memory-less mutations, consisting in drawing the new value of $J_{ij}$ uniformly at random in $[-1,1]$, independently of its previous value. Among other possible choices, we may consider: (i) discrete couplings, taken at random in a finite set of values, $\pm\{0,0.01,0.02,0.05,0.1,0.2,0.5,1\}$; (ii) a sum-rule, where each mutation adds to the current value a normally distributed random variable: $J_{ij} \rightarrow J^*_{ij}=J_{ij} + \mathcal{N}(0,\sigma_s^2)$; (iii) a product-rule, where each mutation multiplies the current value by a Gaussian variable: $J_{ij} \rightarrow J_{ij}^*=J _{ij}\times \mathcal{N}(0,\sigma_p^2)$. We implemented these rules by mapping values $J_{ij}^*>1$ to $J_{ij}^*=1$ and values $J_{ij}^*<-1$ to $J_{ij}^*=-1$, to ensure that the couplings remain bounded. The results are presented in Fig.\ S\ref{fig:mutations}, showing robustness of our conclusions with respect to the mutational process.\\

\begin{figure}[H]
\begin{center}
\includegraphics[width=\linewidth]{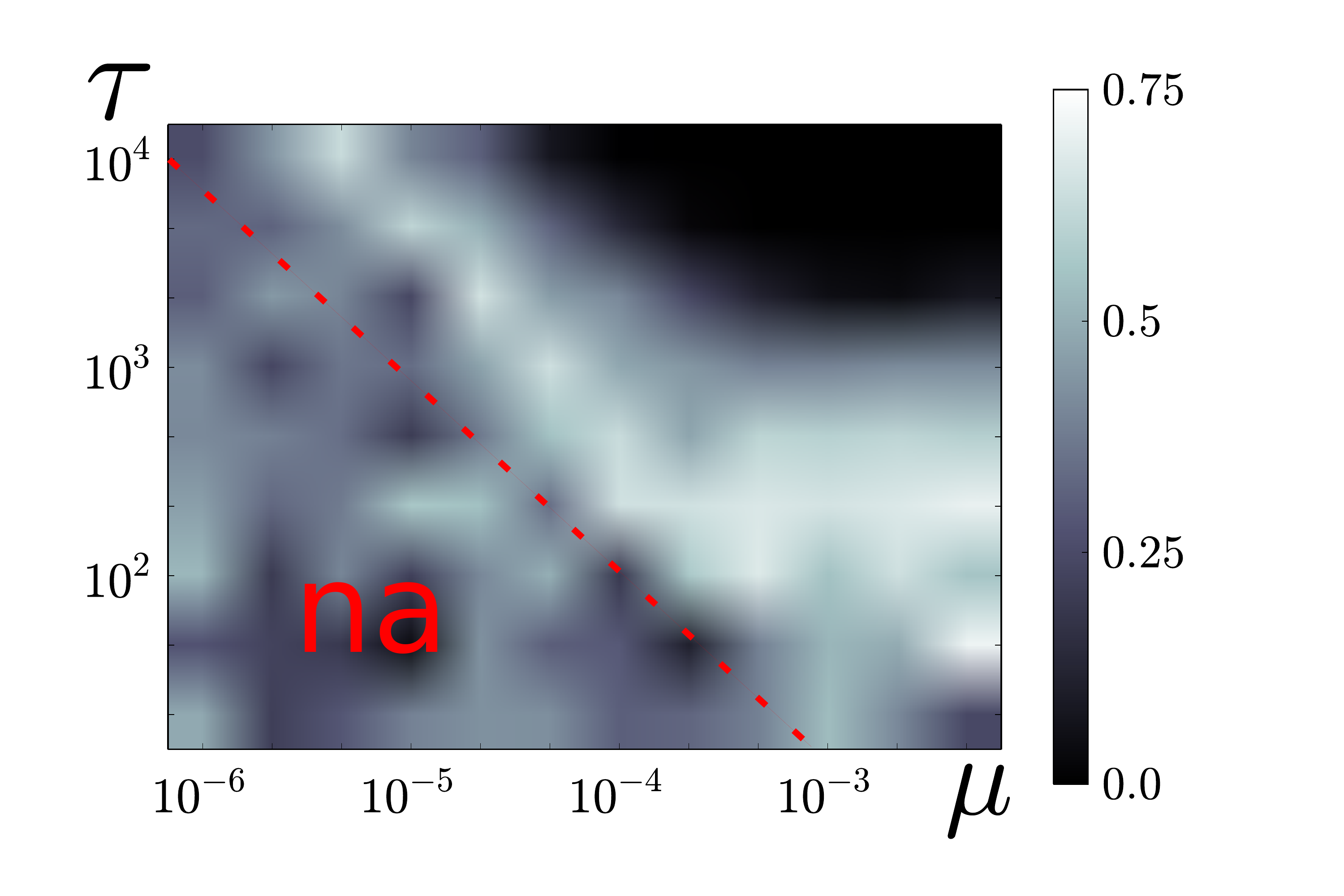}
\caption{Sparsity of evolved systems as a function of $\mu$ and $\tau$ for systems defined on a three-dimensional $5\times 5\times 5$ square lattice with periodic conditions along two dimensions and regulatory and active sites at the two open boundaries of the third. These results generalize those of Fig.\ \ref{fig:parameters} to a three-dimensional system.\label{fig:geometry}}
\end{center}
\end{figure}

{\bf Alternative geometries --}
We presented our results with a two-dimensional square lattice but our model is equally solvable for any geometry. Fig.\ S\ref{fig:geometry} thus shows results obtained with a $5\times 5\times 5$ three-dimensional regular cubic lattice with periodic boundary conditions along two dimensions and the regulatory and active sites defined on the faces associated with the third dimension (a three-dimensional generalization of the cylinder).\\

\begin{figure}[H]
\begin{center}
\includegraphics[width=\linewidth]{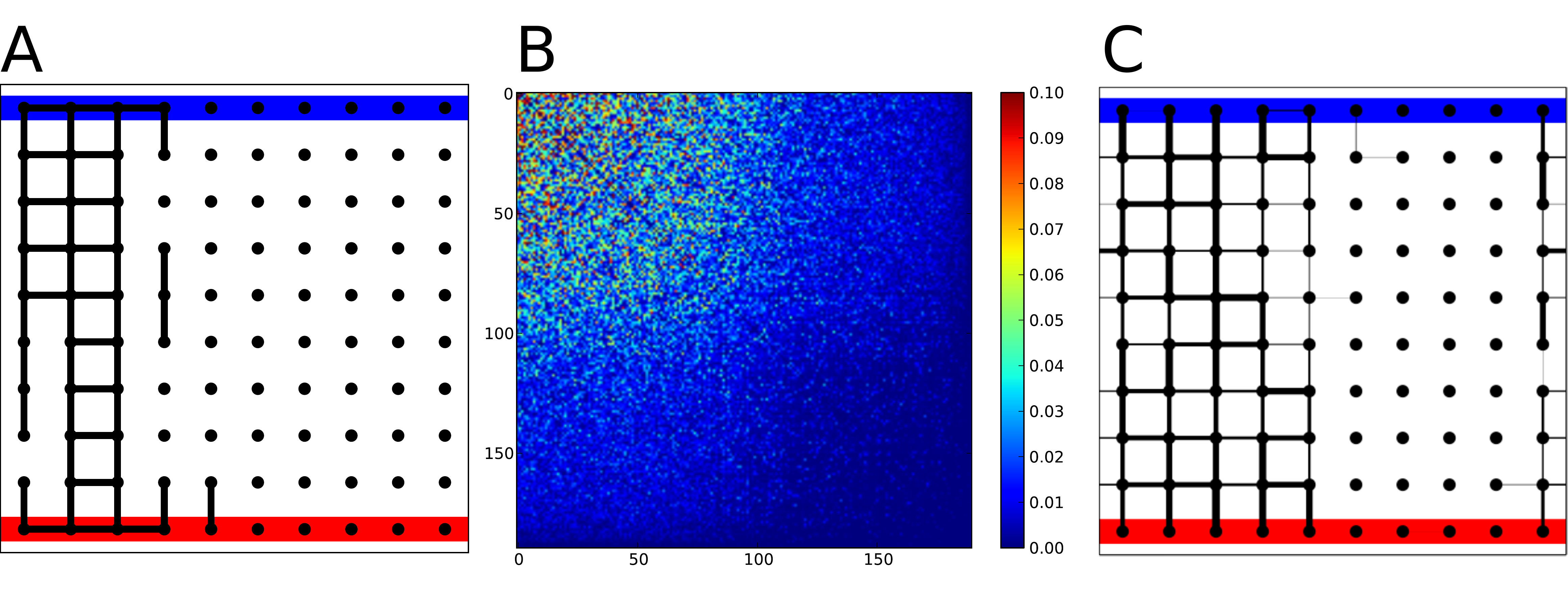}
\caption{Analysis of coevolution. {\bf A.}~Analogous to Fig.\ \ref{fig:grids} with parameters $\tau = 200$ and $\mu = 5.10^{-5}$. {\bf B.}~Matrix of covariations between couplings, obtained by comparing systems generated from independent trajectories originating from a common initial population; the positions are ordered by the principal eigenvector of the matrix. {\bf C.}~Mapping on the structure of the top couplings identified in B, showing a correspondence with the couplings identified in A based on the fitness cost of individual mutations.\label{fig:coevo}}
\end{center}
\end{figure}

{\bf Coevolution --}
To analyze whether the functionally most significant couplings are subject to coevolution, we took a population obtained with $\tau = 200$, $\mu = 5.10^{-5}$, and used it as a common initial condition for $100$ independent trajectory subject to the same $\tau,\mu$. After $3\tau=600$ generations, we computed a  matrix of covariance between couplings, $\mathcal{C}_{ij,kl} = |\langle J_{ij}J_{kl}\rangle-\langle J_{ij}\rangle\langle J_{kl}\rangle|$, where $\langle\ldots\rangle$ denote an average over the different populations (the absolute value is taken to treat equivalently positive and negative covariations). We then performed a principal component analysis to identify the couplings that covary the most: Fig.\ \ref{fig:coevo}B shows the matrix $\mathcal{C}$ where the couplings are ordered based on their contribution to the principal eigenvector. The top positions defined by this principal component define a sector (Fig.\ \ref{fig:coevo}C), which overlaps with the sector defined, as in Fig.\ \ref{fig:grids}B, based on the fitness cost of punctual mutations (Fig.\ \ref{fig:coevo}A).\\

\begin{figure}[H]
\begin{center}
\includegraphics[width=.9\linewidth]{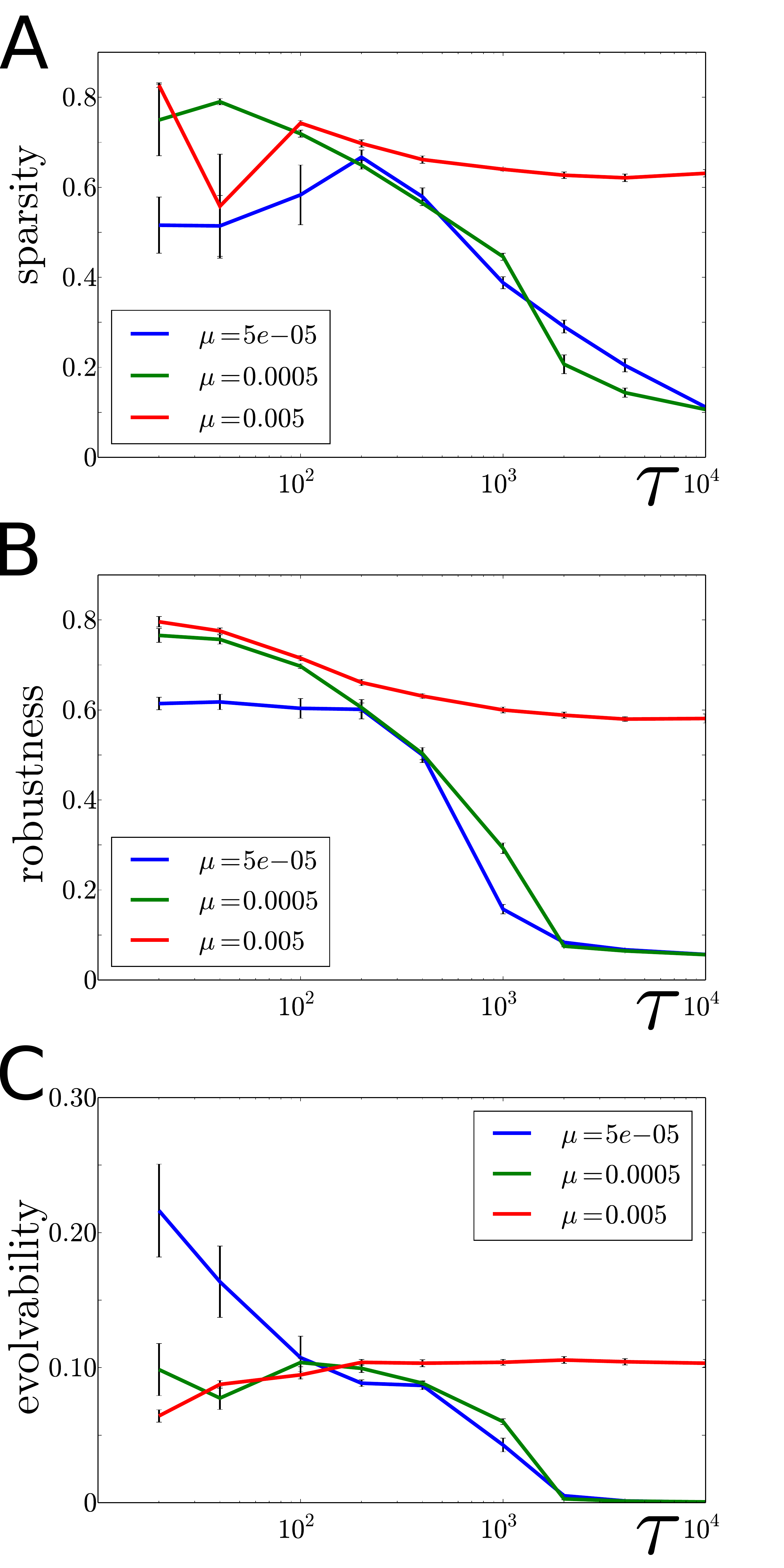}
\caption{Sparsity, robustness and evolvability of evolved systems as a function of the period $\tau$ of environmental changes. The error-bars are standard deviations over 10 simulations. \label{fig:slices_tau}}
\end{center}
\end{figure}

\begin{figure}[H]
\begin{center}
\includegraphics[width=.9\linewidth]{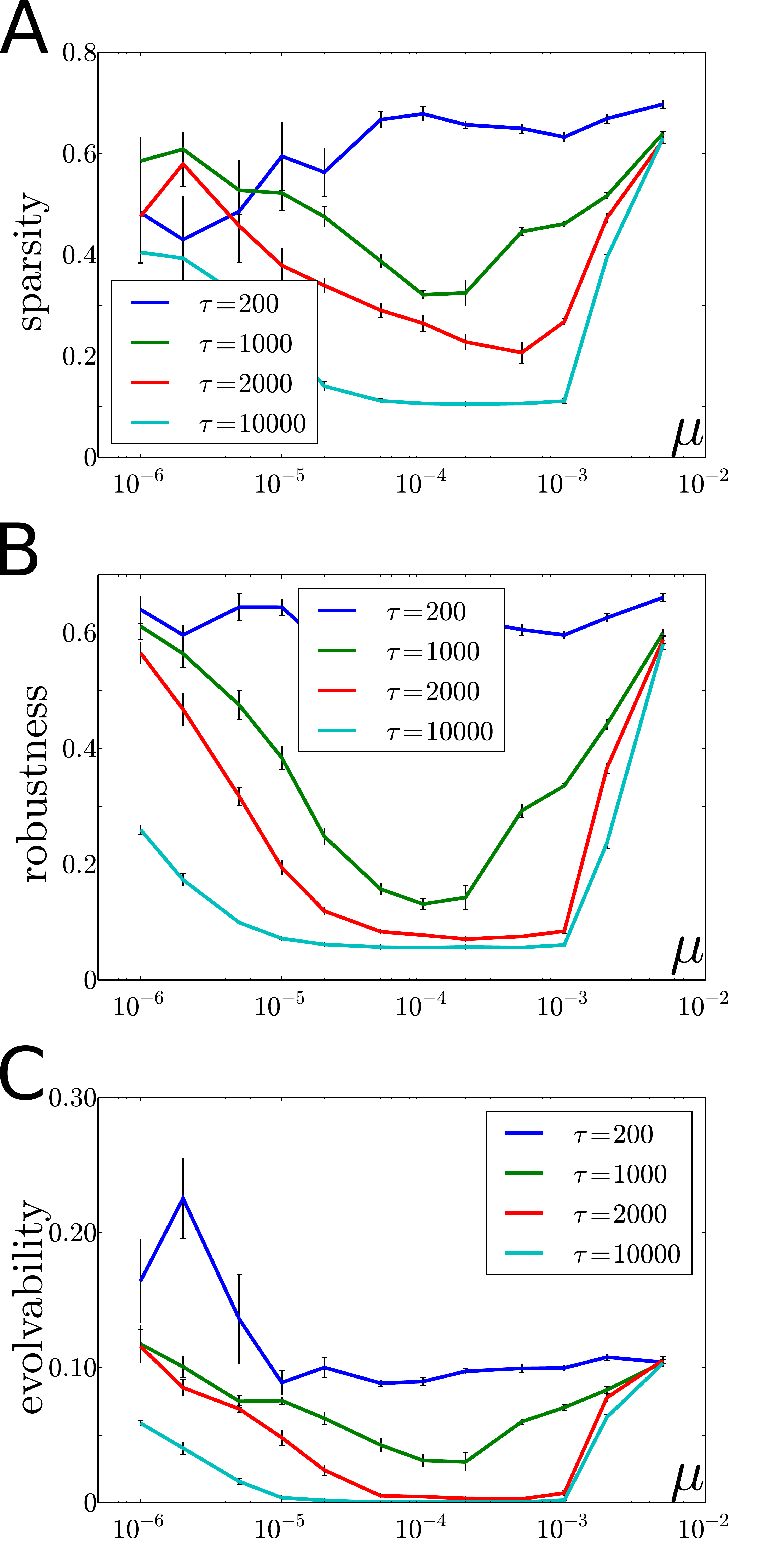}
\caption{Sparsity, robustness and evolvability of evolved systems as a function of the mutation rate $\mu$. The error-bars are standard deviations over 10 simulations. \label{fig:slices_mu}}
\end{center}
\end{figure}

{\bf Robustness -- } The robustness of a system, shown in Figs.\ S\ref{fig:slices_tau}B-S\ref{fig:slices_mu}B, is defined as the fraction of mutations that do not cause a significant fitness cost. Formally,
\beq
R(J|h)=\left\langle\theta\left(\delta\phi_{ij}^*< 0.01 \right)\right\rangle_{ij,*}
\eeq
where $\theta(x)=1$ if $x\geq 0$ and $0$ otherwise, and where $\langle.\rangle_{ij,*}$ is an average over the pairs $ij$ and over the possible values of $J^*_{ij}$.\\

{\bf Evolvability -- } The evolvability of a system, shown in Figs.\ S\ref{fig:slices_tau}C-S\ref{fig:slices_mu}C, is defined as the fraction of mutations that cause a significant advantage when the selective pressure changes. Formally,
\beq
E(J|h')=\left\langle\theta\left(\delta\phi_{ij}^*>0.2 \right)\right\rangle_{ij,*}.
\eeq
It differs from the definition of $R$ by the field $h'$ which is distinct from the field $h$ in which the system most recently evolved. When considering an environment alternating periodically between two values $h^{(1)}$ and $h^{(2)}$, we thus take the systems at the end of a period of constant selective pressure under $h^{(1)}$ and define robustness as $R(J|h^{(1)})$ and evolvability as $E(J|h^{(2)})$.

\begin{figure}[H]
\begin{center}
\includegraphics[width=\linewidth]{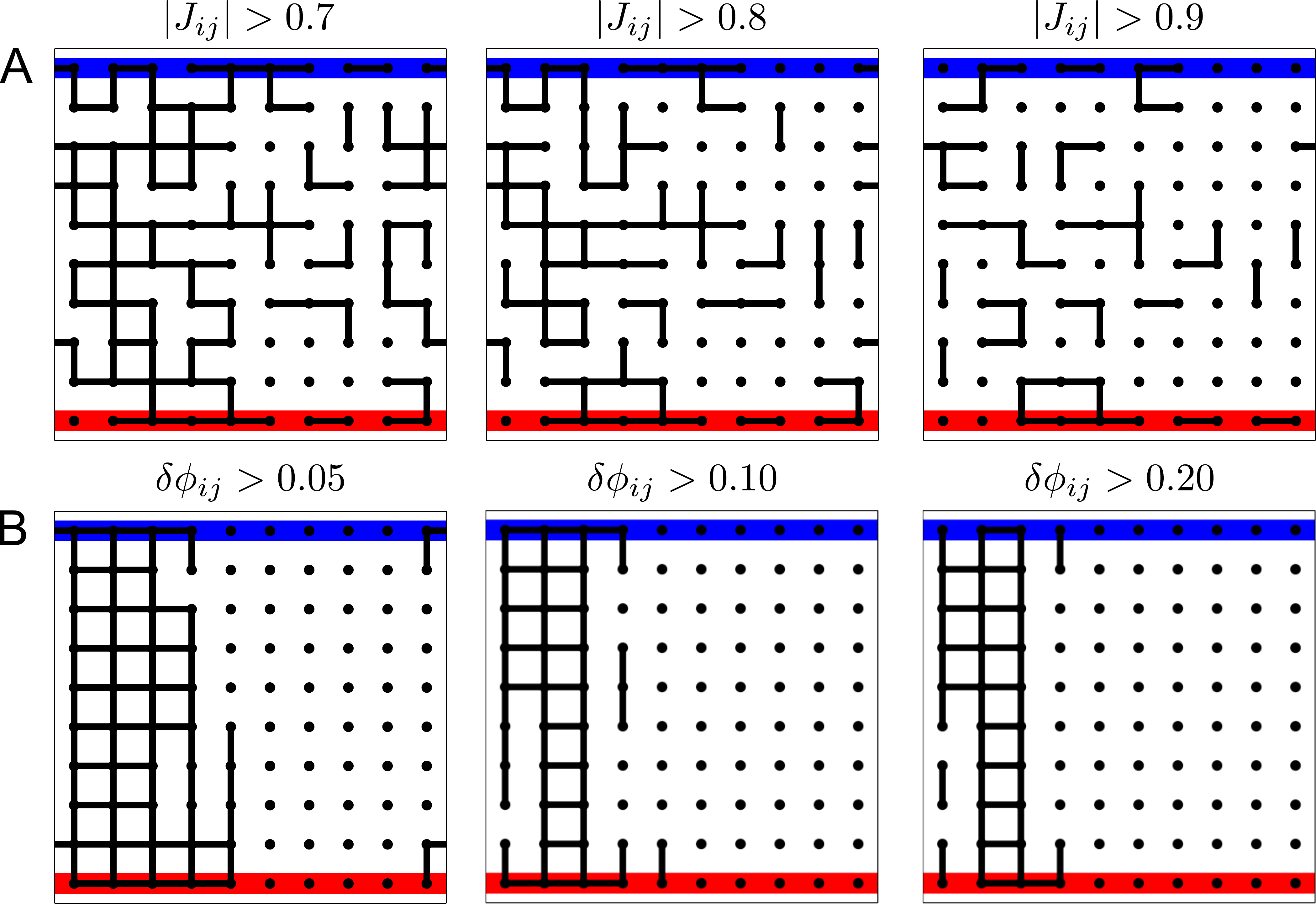}
\caption{For the system associated with $\tau=200$ in Fig.\ \ref{fig:grids}, couplings $|J_{ij}|$ and functional constraints $\delta\phi_{ij}$ above different cut-offs (Fig.\ \ref{fig:grids} corresponds to $|J_{ij}|>0.8$ and $\delta\phi_{ij}>0.1$). \label{fig:cutoffs}}
\end{center}
\end{figure}

\begin{figure}[H]
\begin{center}
\includegraphics[width=\linewidth]{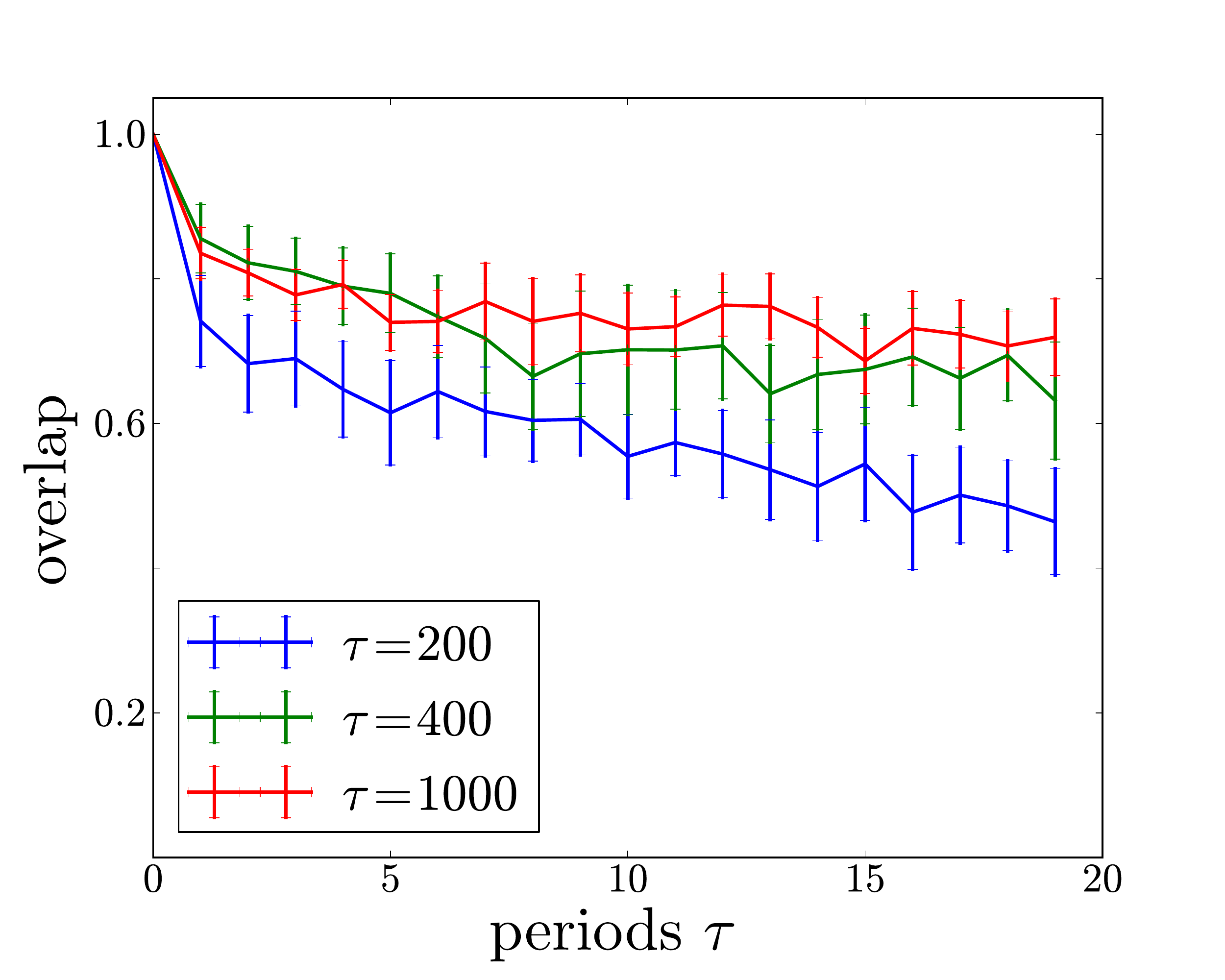}
\caption{Overlap of functionally significant couplings ($\delta\phi_{ij}>0.1$) between a system at time $t_0$ and a system at time $t_0+t$ along a same evolutionary trajectory as a function of time $t$, counted in number of periods $\tau$ (the error-bars are standard deviations over 100 simulations).
\label{fig:stability}}
\end{center}
\end{figure}

\begin{figure}[H]
\begin{center}
\includegraphics[width=\linewidth]{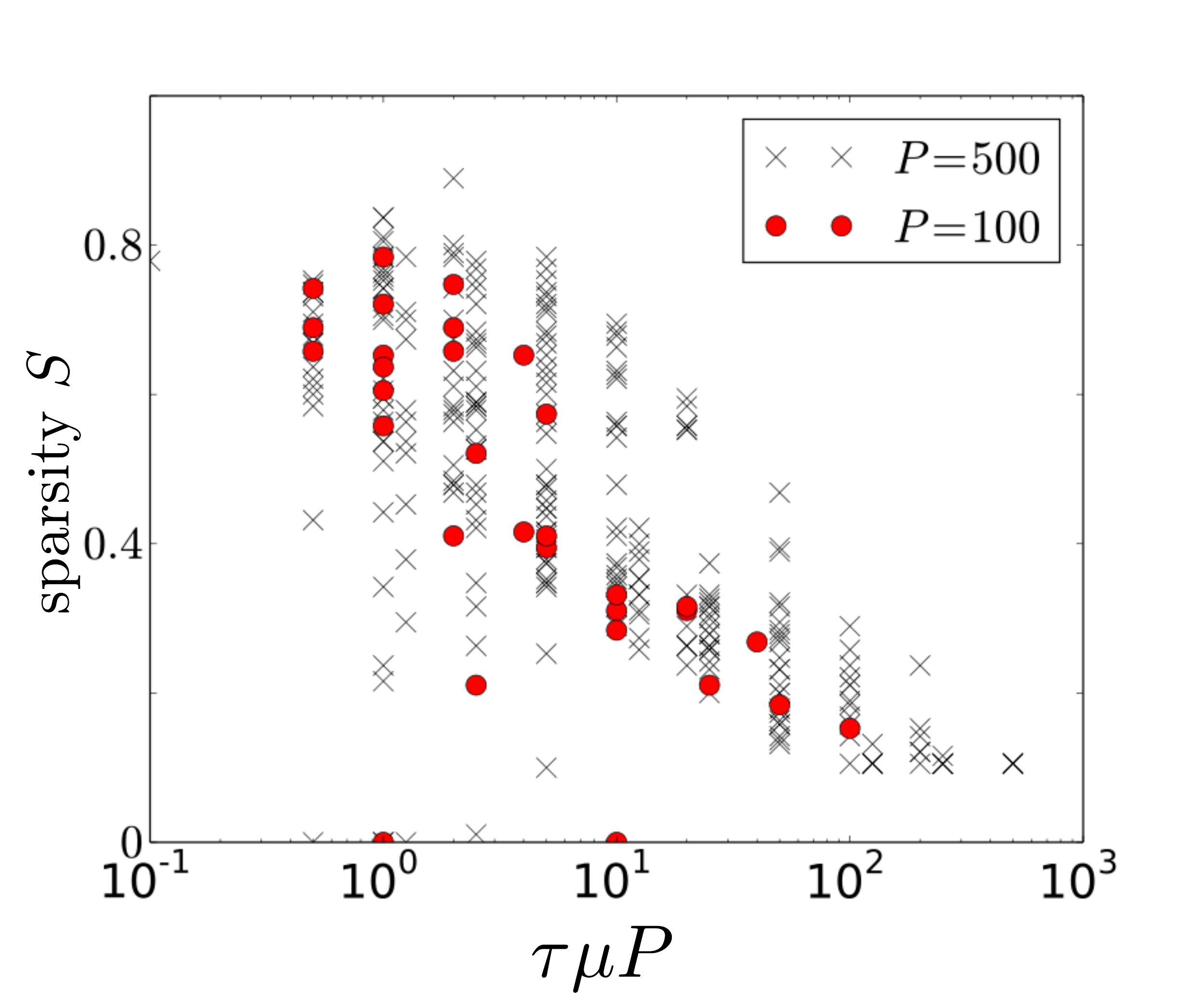}
\caption{Sparsity as a function of the scaling variable $\tau\mu P$ for systems obtained from evolutionary dynamics with different values of the period $\tau$ of environmental changes and mutation rate $\mu$ (in the range of values used for Fig.\ \ref{fig:parameters}) and for two population sizes, $P=100$, $500$. \label{fig:scaling}}
\end{center}
\end{figure}

\begin{figure}[H]
\begin{center}
\includegraphics[width=\linewidth]{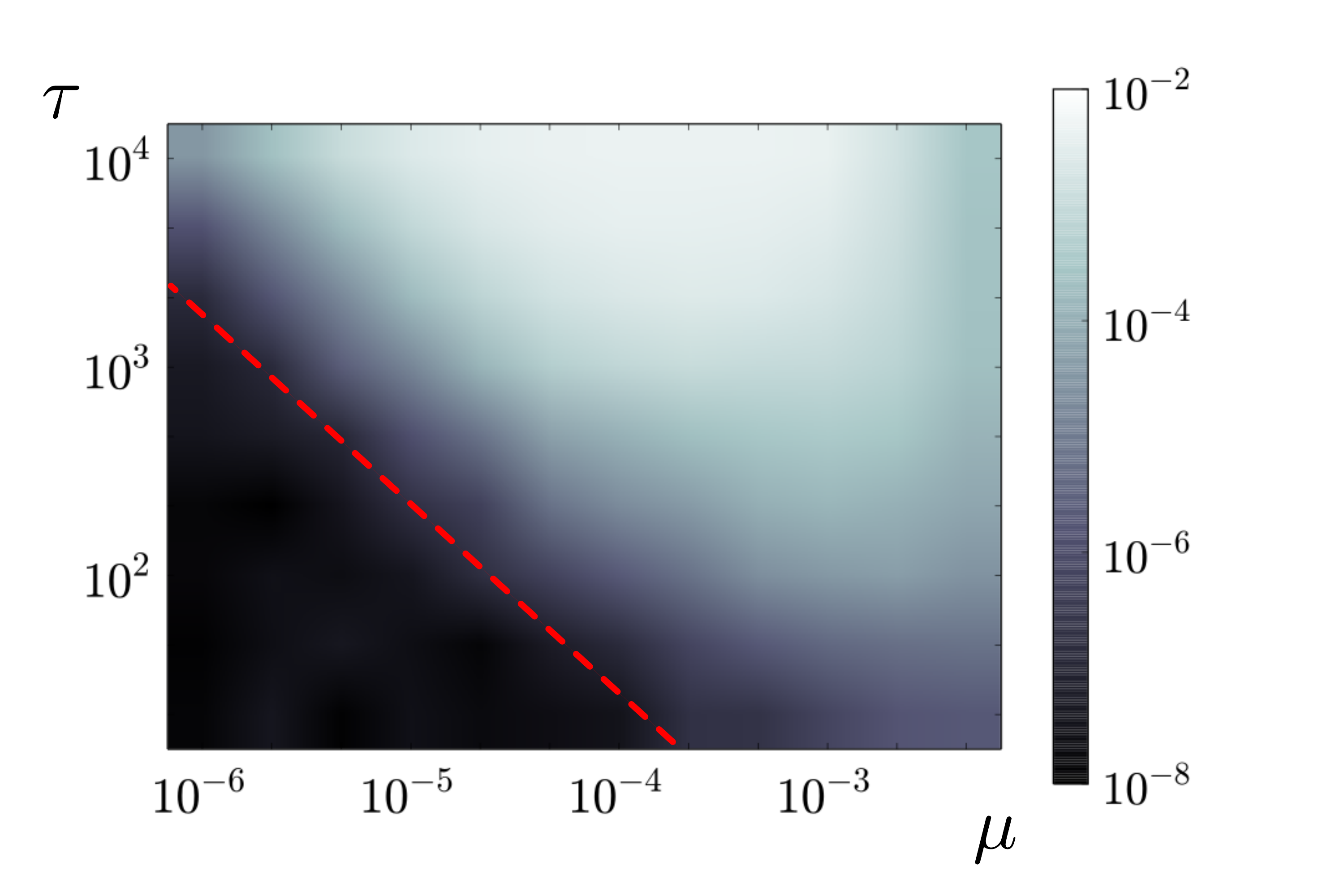}
\caption{Fitness of the population as a function od $\mu$ and $\tau$, the axis and the red line are the same as in Fig.\ \ref{fig:parameters}. The red line ($\phi=10^{-7}$) corresponds to the typical maximal value of allosteric efficiency in populations of $P=500$ random systems. Below this line, the populations may be considered as non-adapted.\label{fig:unfit}}
\end{center}
\end{figure}

\end{document}